\documentclass[pre,twocolumn,amsmath,amssymb]{revtex4}

\usepackage{graphicx}

\newcommand{\vecs}{\mathbf{s}}
\newcommand{\vecx}{\mathbf{\hat{x}}}
\newcommand{\vecy}{\mathbf{\hat{y}}}
\newcommand{\vecz}{\mathbf{\hat{z}}}
\newcommand{\vecxx}{\mathbf{x}}
\newcommand{\vecg}{\mathbf{g}}

\begin{document}

\title{Algorithms for Brownian first passage time estimation}

\author{Artur B. Adib}
\email{adiba@mail.nih.gov}
\affiliation{
Laboratory of Chemical Physics, NIDDK, National Institutes of Health, Bethesda, Maryland 20892-0520, USA
}

\date{\today}

\begin{abstract}
A class of algorithms in discrete space and continuous time for Brownian first passage time estimation is considered. A simple algorithm is derived that yields exact mean first passage times (MFPT) for linear potentials in one dimension, regardless of the lattice spacing. When applied to nonlinear potentials and/or higher spatial dimensions, numerical evidence suggests that this algorithm yields MFPT estimates that either outperform or rival Langevin-based (discrete time, continuous space) estimates.
\end{abstract}

\maketitle

Brownian dynamics is one of the most widespread models of temporal evolution for systems displaying stochastic behavior \cite{mazo-book}. Its popularity stems no doubt in part from its simplicity, which allows one to carry out analytical work to great lengths, but also from its generality, as many dynamical systems ranging from biological molecules \cite{berg-book} to financial markets \cite{bouchaud-book} are often well approximated by this model.

In general, however, the solution of most Brownian problems is not known in closed analytical form, requiring one to resort to numerical simulations. Almost invariably, such solutions are obtained by discretizing the Langevin equation in time, and iterating the ensuing difference equation (see e.g. \cite{allen-tildesley87}). Here I propose to discretize space instead, leaving time continuous. There are many ways of going about this procedure, and different algorithms can be obtained depending on the desired context. In this paper I will focus on the design of algorithms suited for the computation of mean first passage times (MFPT) to a given boundary \cite{redner-book}, which plays a particularly important role in theories of chemical kinetics \cite{szabo80,szabo81}.

\begin{figure}
\begin{center}
\includegraphics[width=200pt]{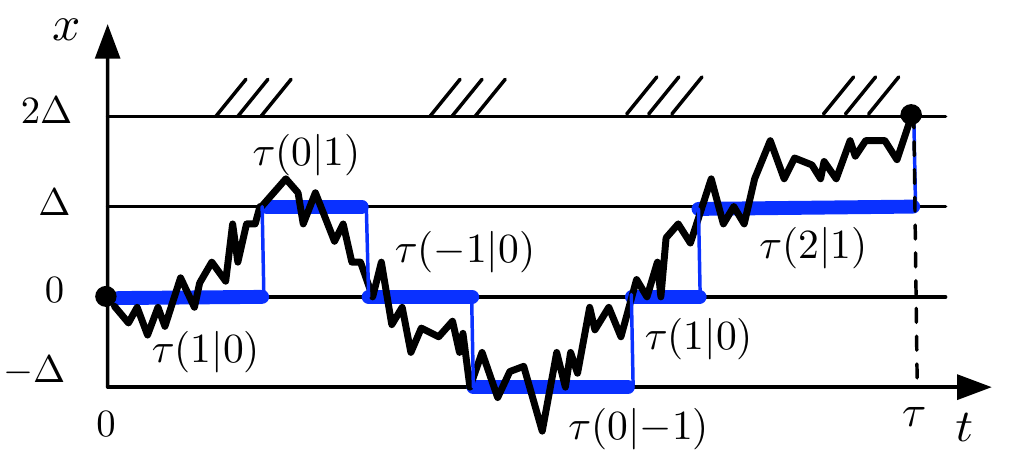} 
\caption{A continuous Brownian trajectory (wiggly black line) and its discrete counterpart (straight blue lines), illustrating a first passage problem from $x=0$ to the absorbing boundary at $x=2 \Delta$. The discrete states are labeled $s = 0, \pm 1, \pm 2$, and etc, corresponding to $x = 0, \pm \Delta, \pm 2 \Delta$, and etc. For both types of trajectories, the total first passage time $\tau$ is the sum of the conditional first passage times $\tau(s_{i+1}|s_i)$ from state $s_i$ to the next state $s_{i+1}$, where ``conditional'' means that the particle did not cross the other adjacent state before crossing $s_{i+1}$ \cite{redner-book}. The discrete trajectories are constructed so that their MFPT $\langle \tau \rangle$ is the same as that of the original Brownian trajectories (see text).}
\label{fig:illustration}
\end{center}
\end{figure}

To introduce the basic idea behind the present algorithm, let us focus on the simple one-dimensional problem shown in Fig.~\ref{fig:illustration}. The illustration depicts a typical Brownian trajectory in a first-passage problem from $x=0$ to $x=2\Delta$. This problem is characterized by an ensemble of continuous trajectories that start from $x=0$ at $t=0$, and cross the absorbing boundary $x=2 \Delta$ only once at some time $t=\tau$; $\tau$ is thus the first passage time of the trajectory. Our goal is to design algorithms that generate discrete trajectories (blue lines in Fig.~\ref{fig:illustration}) that ``hop'' from site to site so that their MFPT to the absorbing boundary approximate that of the original, continuous problem.

The outline of the derivation is as follows. First, the mean first passage time of the continuous Brownian problem will be recast in terms of two quantities defined on an arbitrary lattice, namely conditional mean first passage times and splitting probabilities (Eq.~(\ref{tau-final})). This intermediate result will allow us to design lattice algorithms that reproduce the MFPT of the original Brownian problem by demanding that their conditional MFPTs and splitting probabilities be equal to those of the Brownian problem. As these two quantities are generally not algebraic for nonlinear potentials, they will be evaluated based on a linear approximation in the region delimited by the nearest neighbor sites (Eqs.~(\ref{tau-linear}) and (\ref{phi-linear})). By additionally demanding the sites to be uniformly spaced, this will allow us to write a generic rate equation that can be generalized to higher dimensions (Eqs.~(\ref{rate}) and (\ref{k})). Finally, this rate equation is simulated by standard means, e.g. using Gillespie's algorithm \cite{wilkinson-book}.

Going back to Fig.~\ref{fig:illustration}, we see that the first passage time of any Brownian trajectory can be decomposed as a sum of intermediate times $\tau(s_{i+1}|s_i)$. Thus, the mean first passage time from state $s_1$ to state $s_N$ in the restricted ensemble of trajectories that pass through a given time-ordered sequence of states $s^N = \{ s_1, s_2, \ldots, s_N \}$ is
\begin{equation}
  \langle \tau(s^N) \rangle = \sum_{i=1}^{N-1} \langle \tau(s_{i+1}|s_i) \rangle.
\end{equation}
The quantity $\langle \tau(s_{i+1}|s_i) \rangle$ is the conditional mean first passage time from state $s_i$ to state $s_{i+1}$, where the term ``conditional'' means that the particle is not allowed to pass through the other adjacent state \cite{redner-book} (e.g. $\tau(1|0)$ is the first passage time from $s=0$ to $s=1$, conditional on not passing through $s=-1$). Note that the individual terms of this sum depend only on the present and next states, $s_i$ and $s_{i+1}$ respectively. This is only true for Markovian dynamics, which is assumed to be the case for the present Brownian problem. The total MFPT $\langle \tau \rangle$ is thus obtained by taking the average of $\langle \tau(s^N) \rangle$ over all permissible sequences of states $s^N$, i.e.
\begin{equation}
  \langle \tau \rangle = \sum_{N} \sum_{s^N} p(s^N) \langle \tau(s^N) \rangle,
\end{equation}
where $p(s^N)$ is the probability that the particular sequence of states $s^N$ will be realized, and the double sum is over all possible sequences of states that take the particle from its original position to the absorbing boundary.

Now, for Markovian dynamics, the probability $p(s^N)$ can be decomposed as a product of splitting probabilities $\phi$, where $\phi(s_{i+1}|s_i)$ is the probability that a particle originally at $s_i$ will pass through $s_{i+1}$ before passing through the other adjacent state (e.g. $\phi(2|1)$ is the probability that the particle originally at $s=1$ will pass through $s=2$ before $s=0$). This finally gives the result
\begin{equation} \label{tau-final}
  \langle \tau \rangle = \sum_{N} \sum_{s^N} \left( \prod_{i=1}^{N-1} \phi(s_{i+1}|s_i) \right) \sum_{i=1}^{N-1} \langle \tau(s_{i+1}|s_i) \rangle.
\end{equation}
The main conclusion from this derivation is that the MFPT of our Brownian problem is fully specified by the splitting probabilities and conditional MFPTs defined on an arbitrary lattice (although we have chosen a uniform lattice anticipating the development below, this derivation is clearly more general). It thus follows that any other dynamical system that has the same $\phi(s'|s)$ and $\langle \tau(s'|s) \rangle$ for all adjacent sites $s',s$ as the original Brownian problem also has the same MFPT $\langle \tau \rangle$. In turn, this suggests that the design of MFPT algorithms on a lattice should focus on reproducing as closely as possible these two quantities from the original Brownian problem.

For one dimensional Brownian problems, both $\phi(s\pm 1|s)$ and $\langle \tau(s\pm 1|s) \rangle$ can be reduced to simple quadrature \cite{redner-book}. An additional simplification occurs when the particle is subject to a linear potential and the lattice is uniformly spaced, in which case two things happen: first, the integrals reduce to algebraic expressions, and second the conditional MFPTs $\langle \tau(s+1|s) \rangle$ and $\langle \tau(s-1|s) \rangle$ become coincident and equal to the unconditional mean exit time, $\langle \tau(s) \rangle$. This second observation allows us to write down a rate equation governing the dynamics on the lattice, which can then be generalized to higher dimensions.

To be specific, consider a particle evolving according to the Smoluchowski equation \cite{mazo-book}
\begin{equation} \label{smol}
  \frac{\partial p}{\partial t} = D \nabla^2 p + D \nabla \cdot ( \nabla U \, p),
\end{equation}
and subject to the linear potential $U(x) = \alpha x$, where for simplicity of notation energy is measured in units of $k_B T$. For such one dimensional potentials, the (unconditional) mean first passage time \cite{redner-book,gardiner-book} from $s$ to the adjacent positions $s \pm 1$ is, exactly,
\begin{equation} \label{tau-linear}
  \langle \tau(s) \rangle = \frac{\Delta}{D \alpha} \frac{e^{\alpha \Delta} - 1}{ e^{\alpha \Delta} + 1},
\end{equation}
while the splitting probabilities are
\begin{equation} \label{phi-linear}
  \phi(s \pm 1 | s) = \frac{1}{1 + e^{\pm \alpha \Delta}}.
\end{equation}
Given $\langle \tau(s) \rangle$ and $\phi(s \pm 1 | s)$, a lattice rate equation can be constructed consistent with these quantities. Indeed, consider a kinetic scheme for the states $s-1, s, s + 1$ with outgoing rates from $s$ given by $k(s \pm 1 | s)$. The lifetime $\langle \tau(s) \rangle$ in the state $s$ is then $\langle \tau(s) \rangle^{-1} = k(s - 1 | s) + k(s + 1|s)$, while the splitting probabilities are $\phi(s \pm 1 |s) = \langle \tau(s) \rangle k(s\pm 1|s)$. Solving these equations for the rates and using the results for linear potentials above, we get
\begin{equation} \label{k1d}
  k(s \pm 1 | s) = \pm \frac{D \alpha}{\Delta} \frac{1}{e^{\pm \alpha \Delta} - 1},
\end{equation}
where these rates are to be used in the rate equation
\begin{multline} \label{rate1d}
  \frac{d p(s;t)}{d t} = k(s|s+1) p(s+1;t) + k(s|s-1) p(s-1;t) \\ 
  - [k(s - 1 | s) + k(s + 1|s)] p(s;t).
\end{multline}

\begin{figure}
\begin{center}
\includegraphics[width=120pt]{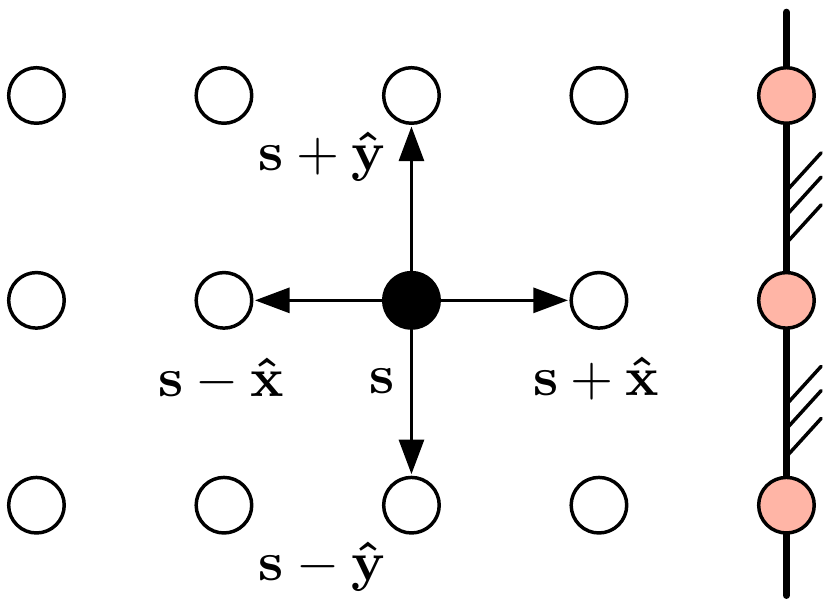} 
\caption{Illustration of a two dimensional implementation of Eqs.~(\ref{rate})-(\ref{k}). The arrows represent the four outgoing rates from $\vecs$, $k(\vecs \pm \vecx | \vecs)$ and $k(\vecs \pm \vecy | \vecs)$. The boundary sites lie along the vertical line on the right and are highlighted in red. The particle ``dies'' whenever it visits one such site.}
\label{fig:2d}
\end{center}
\end{figure}

Equations (\ref{k1d}) and (\ref{rate1d}) form the foundation of the proposed algorithm. For linear potentials in one spatial dimension, the algorithm yields exact MFPTs. For nonlinear potentials, $\alpha$ is to be replaced by the slope of the potential at the position corresponding to site $s$ (local linear approximation). In higher dimensions, the rate equation Eq.~(\ref{rate1d}) can be generalized by taking the rates along each coordinate to be the one dimensional result already derived. Thus, the general form of our rate equation takes the form
\begin{equation} \label{rate}
  \frac{d p(\vecs;t)}{d t} = \sum_{\vecs' = \text{n.n.}} \left[ k(\vecs|\vecs') p(\vecs';t) - k(\vecs'|\vecs) p(\vecs;t) \right],
\end{equation}
where the sum is over the nearest neighbors of $\vecs$, and
\begin{equation} \label{k}
  k(\vecs \pm \vecz | \vecs) = \pm \frac{D U_z(\vecs)}{\Delta} \frac{1}{e^{\pm U_z(\vecs) \Delta} - 1}.
\end{equation}
In this last equation, $\vecz$ is a unit basis vector along any of the coordinates, and $U_z(\vecs)$ is the partial derivative of the potential with respect to that coordinate evaluated at the position corresponding to the site $\vecs$. For simplicity, cartesian coordinates and square lattices are being assumed (see Fig.~\ref{fig:2d}).

Before discussing how to simulate the above rate equation, let us check that in the continuum limit we are exactly solving the Smoluchowski equation (Eq.~(\ref{smol})). When $\Delta$ is small in Eq.~(\ref{k}), we have to leading order
\begin{equation}
  k(\vecs \pm \vecx | \vecs) = \frac{D}{\Delta^2} \left( 1 \mp \frac{U_x(\vecs) \Delta }{2} + \ldots \right).
\end{equation}
Substituting these rates into Eq.~(\ref{rate}) and mapping finite differences into differential operators, we indeed obtain Eq.~(\ref{smol}). This shows that, although our method was designed with MFPT estimation in mind, the ensuing algorithm actually generates exact trajectories in the continuum limit, much like the Langevin algorithm becomes exact when the time step goes to zero.

\begin{figure}
\begin{center}
\includegraphics[width=190pt]{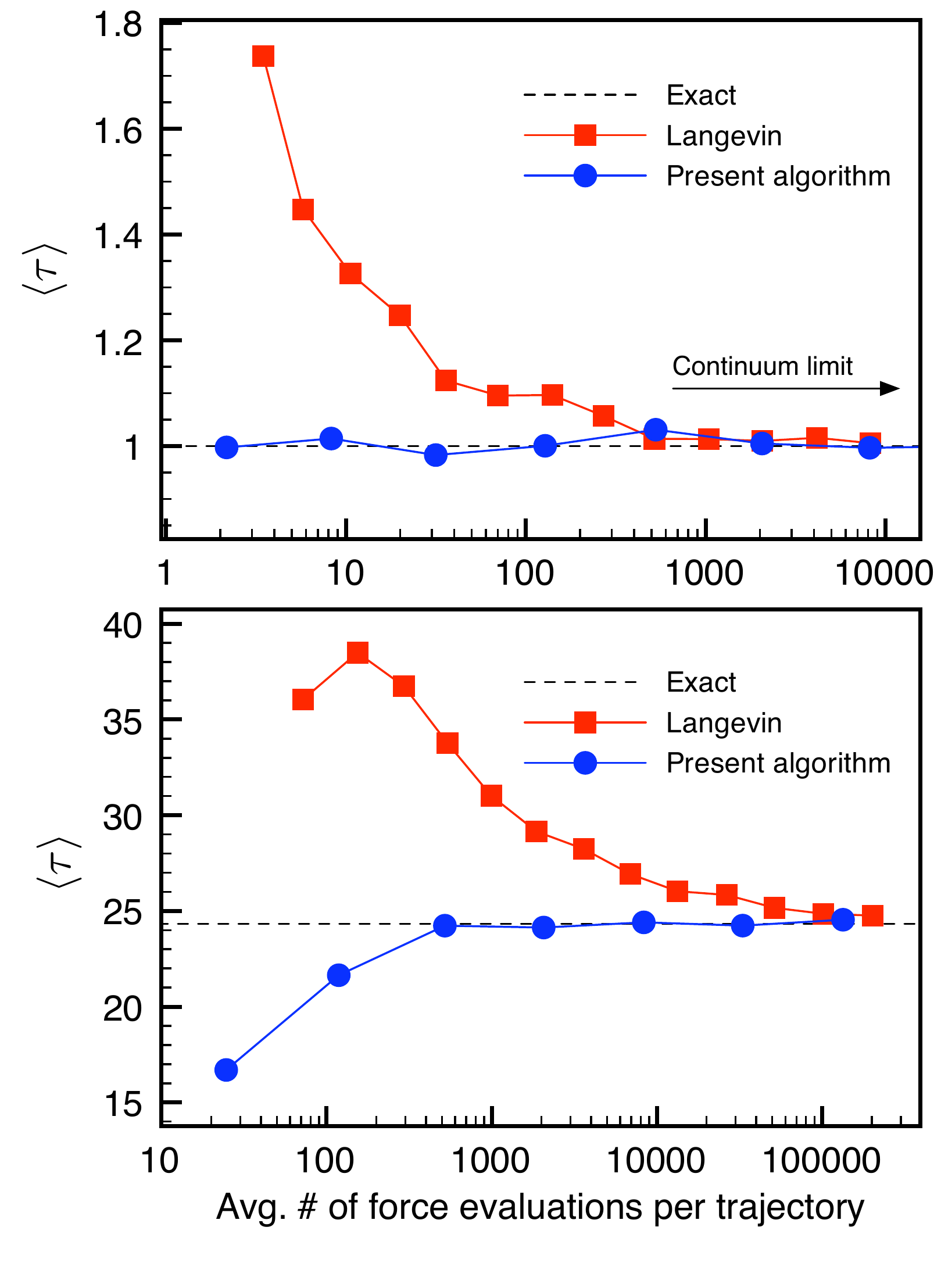} 
\caption{Numerical results in one dimension, comparing Langevin's algorithm (Eq.~(\ref{langevin}), red squares) with the present algorithm (Eqs.~(\ref{k1d})-(\ref{rate1d}), blue circles). {\em Top:} Mean first passage time from $x=0$ to $x=1$ for the linear potential $U(x) = -x$. The exact result obtained by analytical integration \cite{gardiner-book} is $\langle \tau \rangle = 1$ (dashed line). {\em Bottom:} MFPT from $x=0$ to $x=\sqrt{6}$ for the harmonic potential $U(x) = x^2/2$. The ``exact'' result obtained by numeric quadrature \cite{gardiner-book} is $\langle \tau \rangle = 24.324$ (dashed line). For both problems $D=1$. The error bars are of the size of the symbols and hence not shown. For the Langevin algorithm, the symbols correspond to decreasing values of the time step $\Delta t$, from left to right (e.g. $\Delta t = 0.5, 0.25, 0.125$, etc). For the present algorithm, the symbols correspond to increasing number $n$ of lattice points between the starting point and the boundary, from left to right (e.g. $n = 0, 1, 2, 3$, etc). The average number of force evaluations per trajectory corresponds to the total number of calls to the function $U_x(x)$ divided by the total number of trajectories generated ($10^4$). }
\label{fig:1dresults}
\end{center}
\end{figure}

The simulation of Eqs.~(\ref{rate}) and (\ref{k}) can be performed by means of Gillespie's celebrated algorithm \cite{wilkinson-book}. According to this method, one starts in a given state $\vecs$ and draws an exponentially distributed random number $t$ with mean equal to the reciprocal of the sum of the outgoing rates, i.e. $\langle t \rangle^{-1} = \sum_{\vecs' = n.n.}k(\vecs'|\vecs)$. This is the lifetime of the particle in the state $\vecs$. A decision is then made as to which site among the nearest neighboring states the particle is going next. This is done by assigning the statistical weight $w(\vecs') = k(\vecs'|\vecs) / \sum_{\vecs' = n.n.} k(\vecs'|\vecs)$ to each neighboring site $\vecs'$, and choosing one such site with probability $w(\vecs')$. The particle then moves to this chosen site, and the procedure above is repeated until the particle reaches a boundary site. The sum of the times $t$ until this criterion is satisfied is then the first passage time to the boundary.

\begin{figure}
\begin{center}
\includegraphics[width=190pt]{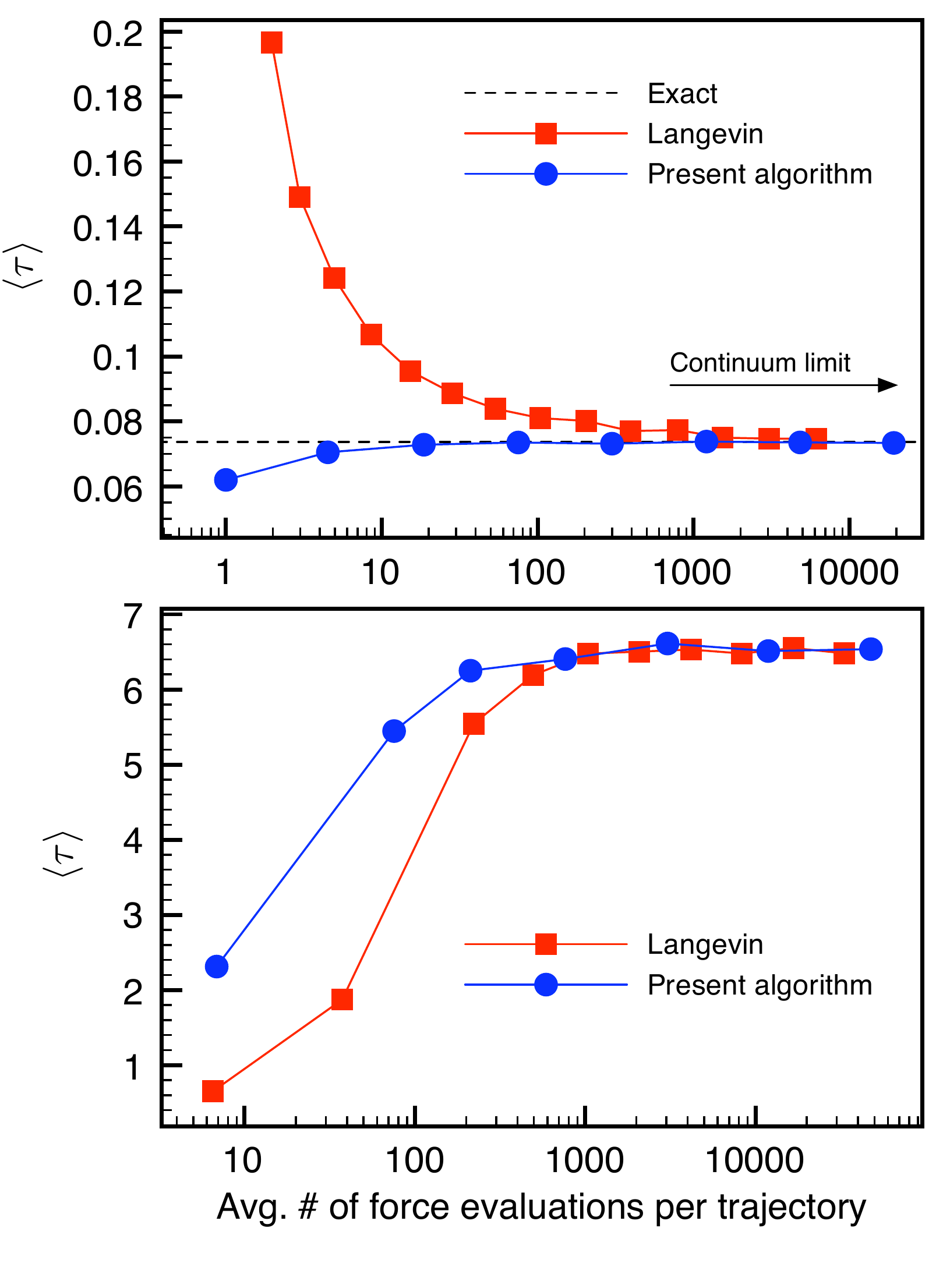} 
\caption{Numerical results in two dimensions. Legends and parameters are the same as in Figure~\ref{fig:1dresults}. {\em Top:} Escape time from the square with corners $(0,0), (1,0), (1,0), (1,1)$, for a free particle starting at $(0.5,0.5)$. The numerically exact result is $\langle \tau \rangle = 0.0736714$ \cite{gardiner-book}. {\em Bottom:} MFPT to the boundary $x=0.5$ for the symmetric double well potential $U(x,y) = [\sqrt{3}(x-1)^2 + y^2]\times[\sqrt{3}(x+1)^2 + y^2]$. The particle starts at the left minimum $(-1,0)$; the other minimum is at $(1,0)$.}
\label{fig:2dresults}
\end{center}
\end{figure}

The above algorithm has been tested on model problems in one and two spatial dimensions (Figures~\ref{fig:1dresults} and \ref{fig:2dresults}, respectively). For comparison, the overdamped Langevin algorithm was also simulated \cite{allen-tildesley87}
\begin{equation} \label{langevin}
  \vecxx(t+\Delta t) = \vecxx(t) - D \Delta t \,  \nabla U(\vecxx(t)) + \sqrt{2 D \Delta t} \, \vecg.
\end{equation}
Here $\Delta t$ is the time step, and $\vecg$ is a Gaussian random vector of zero mean and unit variance. Whenever available, numerically exact results are also reported to illustrate the correctness of the algorithm in the continuum limit. As force computation is the main bottleneck in most simulations, the main figure of merit considered was the average number of force evaluations per trajectory required to achieve a given accuracy level. In all test cases considered, the present algorithm requires considerably fewer force evaluations than the Langevin algorithm, although further experimentation is called for in order to make more general conclusions. 

The algorithm considered above is only one of various strategies that can be used based on Eq.~(\ref{tau-final}). A straightforward improvement is to evaluate the conditional MFPTs explicitly for piecewise linear potentials, so that the resulting algorithm would be exact for such potentials (as opposed to being exact for linear potentials only). This too would result in algebraic expressions for $\langle \tau(\vecs'|\vecs) \rangle$; however its generalization to higher dimensions would be non-trivial, as in this case no simple rate equation can be written (rate constants imply conditional mean lifetimes are the same regardless of the outgoing site). Another possibility is to calculate splitting probabilities and conditional MFPTs for surfaces (instead of lattice points) surrounding the particle. The advantage in this case is that both space and time are treated continuously, and the algorithm applies to any number of dimensions (see e.g. \cite{given97,opplestrup06} for free-particle implementations). The main difficulty here, however, is to find reasonable approximations to such quantities when $U(\vecxx) \neq 0$; to this author's knowledge, such algorithms are yet to be designed.

In summary, in this contribution a new class of algorithms for the estimation of mean first passage times in Brownian dynamics was introduced. In contrast to traditional discrete-time (Langevin) methods, these algorithms treat time continuously and space discretely. Perhaps their most distinguishing feature is that they can yield exact MFPTs regardless of the lattice spacing in some particular cases; for example, the algorithm considered above yields exact MFPTs for linear potentials in one dimension. Numerical results also suggest that the algorithm outperforms Langevin-based estimates in two dimensions. Its efficiency in higher dimensions and/or more complex geometries is currently under investigation.

The author would like to thank Attila Szabo for fruitful comments and suggestions. This research was supported by the Intramural Research Program of the NIH, NIDDK.


\begin{thebibliography}{11}
\expandafter\ifx\csname natexlab\endcsname\relax\def\natexlab#1{#1}\fi
\expandafter\ifx\csname bibnamefont\endcsname\relax
  \def\bibnamefont#1{#1}\fi
\expandafter\ifx\csname bibfnamefont\endcsname\relax
  \def\bibfnamefont#1{#1}\fi
\expandafter\ifx\csname citenamefont\endcsname\relax
  \def\citenamefont#1{#1}\fi
\expandafter\ifx\csname url\endcsname\relax
  \def\url#1{\texttt{#1}}\fi
\expandafter\ifx\csname urlprefix\endcsname\relax\def\urlprefix{URL }\fi
\providecommand{\bibinfo}[2]{#2}
\providecommand{\eprint}[2][]{\url{#2}}

\bibitem[{\citenamefont{Mazo}(2002)}]{mazo-book}
\bibinfo{author}{\bibfnamefont{R.~M.} \bibnamefont{Mazo}},
  \emph{\bibinfo{title}{Brownian Motion: Fluctuations, Dynamics and
  Applications}} (\bibinfo{publisher}{Oxford Univ. Press},
  \bibinfo{address}{Oxford}, \bibinfo{year}{2002}).

\bibitem[{\citenamefont{Berg}(1993)}]{berg-book}
\bibinfo{author}{\bibfnamefont{H.~C.} \bibnamefont{Berg}},
  \emph{\bibinfo{title}{Random Walks in Biology}}
  (\bibinfo{publisher}{Princeton Univ. Press}, \bibinfo{address}{Princeton},
  \bibinfo{year}{1993}).

\bibitem[{\citenamefont{Bouchaud and Potters}(2003)}]{bouchaud-book}
\bibinfo{author}{\bibfnamefont{J.-P.} \bibnamefont{Bouchaud}} \bibnamefont{and}
  \bibinfo{author}{\bibfnamefont{M.}~\bibnamefont{Potters}},
  \emph{\bibinfo{title}{Theory of Financial Risk and Derivative Pricing: From
  Statistical Physics to Risk Management}} (\bibinfo{publisher}{Cambridge Univ.
  Press}, \bibinfo{address}{Cambridge}, \bibinfo{year}{2003}).

\bibitem[{\citenamefont{Allen and Tildesley}(1987)}]{allen-tildesley87}
\bibinfo{author}{\bibfnamefont{M.~P.} \bibnamefont{Allen}} \bibnamefont{and}
  \bibinfo{author}{\bibfnamefont{D.~J.} \bibnamefont{Tildesley}},
  \emph{\bibinfo{title}{Computer Simulation of Liquids}}
  (\bibinfo{publisher}{Clarendon Press}, \bibinfo{address}{Oxford},
  \bibinfo{year}{1987}).

\bibitem[{\citenamefont{Redner}(2001)}]{redner-book}
\bibinfo{author}{\bibfnamefont{S.}~\bibnamefont{Redner}},
  \emph{\bibinfo{title}{A Guide to First-Passage Processes}}
  (\bibinfo{publisher}{Cambridge Univ. Press}, \bibinfo{address}{Cambridge},
  \bibinfo{year}{2001}).

\bibitem[{\citenamefont{Szabo et~al.}(1980)\citenamefont{Szabo, Schulten, and
  Schulten}}]{szabo80}
\bibinfo{author}{\bibfnamefont{A.}~\bibnamefont{Szabo}},
  \bibinfo{author}{\bibfnamefont{K.}~\bibnamefont{Schulten}}, \bibnamefont{and}
  \bibinfo{author}{\bibfnamefont{Z.}~\bibnamefont{Schulten}},
  \bibinfo{journal}{J. Chem. Phys.} \textbf{\bibinfo{volume}{72}},
  \bibinfo{pages}{4350} (\bibinfo{year}{1980}).

\bibitem[{\citenamefont{Schulten et~al.}(1981)\citenamefont{Schulten, Schulten,
  and Szabo}}]{szabo81}
\bibinfo{author}{\bibfnamefont{K.}~\bibnamefont{Schulten}},
  \bibinfo{author}{\bibfnamefont{Z.}~\bibnamefont{Schulten}}, \bibnamefont{and}
  \bibinfo{author}{\bibfnamefont{A.}~\bibnamefont{Szabo}}, \bibinfo{journal}{J.
  Chem. Phys.} \textbf{\bibinfo{volume}{74}}, \bibinfo{pages}{4426}
  (\bibinfo{year}{1981}).

\bibitem[{\citenamefont{Wilkinson}(2006)}]{wilkinson-book}
\bibinfo{author}{\bibfnamefont{D.~J.} \bibnamefont{Wilkinson}},
  \emph{\bibinfo{title}{Stochastic Modelling for Systems Biology}}
  (\bibinfo{publisher}{Chapman \& Hall/CRC}, \bibinfo{address}{Boca Raton},
  \bibinfo{year}{2006}).

\bibitem[{\citenamefont{Gardiner}(2004)}]{gardiner-book}
\bibinfo{author}{\bibfnamefont{C.~W.} \bibnamefont{Gardiner}},
  \emph{\bibinfo{title}{Handbook of Stochastic Methods for Physics, Chemistry,
  and the Natural Sciences}} (\bibinfo{publisher}{Springer},
  \bibinfo{address}{Berlin}, \bibinfo{year}{2004}), \bibinfo{edition}{3rd} ed.

\bibitem[{\citenamefont{Given et~al.}(1997)\citenamefont{Given, Hubbard, and
  Douglas}}]{given97}
\bibinfo{author}{\bibfnamefont{J.~A.} \bibnamefont{Given}},
  \bibinfo{author}{\bibfnamefont{J.~B.} \bibnamefont{Hubbard}},
  \bibnamefont{and} \bibinfo{author}{\bibfnamefont{J.~F.}
  \bibnamefont{Douglas}}, \bibinfo{journal}{J. Chem. Phys.}
  \textbf{\bibinfo{volume}{106}}, \bibinfo{pages}{3761} (\bibinfo{year}{1997}).

\bibitem[{\citenamefont{Opplestrup et~al.}(2006)\citenamefont{Opplestrup,
  Bulatov, Gilmer, Kalos, and Sadigh}}]{opplestrup06}
\bibinfo{author}{\bibfnamefont{T.}~\bibnamefont{Opplestrup}},
  \bibinfo{author}{\bibfnamefont{V.~V.} \bibnamefont{Bulatov}},
  \bibinfo{author}{\bibfnamefont{G.~H.} \bibnamefont{Gilmer}},
  \bibinfo{author}{\bibfnamefont{M.~H.} \bibnamefont{Kalos}}, \bibnamefont{and}
  \bibinfo{author}{\bibfnamefont{B.}~\bibnamefont{Sadigh}},
  \bibinfo{journal}{Phys. Rev. Lett.} \textbf{\bibinfo{volume}{97}},
  \bibinfo{pages}{230602} (\bibinfo{year}{2006}).

\end{thebibliography}
\end{document}